\documentclass{CHEP2004}

\usepackage{graphicx}

\begin{document}

\title{HEP@HOME - A distributed computing system based on BOINC}

\author{Ant\'onio Amorim\thanks{antonio.amorim@fisica.fc.ul.pt}, Dep. of Physics, Faculty of Science, University of Lisbon\\
Jaime Villate\thanks{villate@fe.up.pt}, Pedro Andrade\thanks{pma@fe.up.pt}, Dep. of Physics, Faculty of Engineering, University of Porto}

\maketitle

%<<< SECTION 0 >>>

\begin{abstract}

Project SETI@HOME has proven to be one of the biggest successes of distributed computing during the last years. With a quite simple approach SETI manages to process large volumes of data using a vast amount of distributed computer power.

To extend the generic usage of this kind of distributed computing tools, BOINC is being developed. In this paper we propose HEP@HOME, a BOINC version tailored to the specific requirements of the High Energy Physics (HEP) community.

The HEP@HOME will be able to process large amounts of data using virtually unlimited computing power, as BOINC does, and it should be able to work according to HEP specifications. In HEP the amounts of data to be analyzed or reconstructed are of central importance. Therefore, one of the design principles of this tool is to avoid data transfer. This will allow scientists to run their analysis applications and taking advantage of a large number of CPUs. This tool also satisfies other important requirements in HEP, namely, security, fault-tolerance and monitoring.

\end{abstract}

%<<< SECTION 1 >>>
\section{INTRODUCTION}

A vast number of scientific applications are increasingly requiring the computation of large amounts of data. The HEP area is one of the best examples of these heavy needs. This diverse demand has contributed to the proliferation of computing and storage systems, thus making computers an integral part of several Grid environments.

In the Large Hadron Collider (LHC) accelerator at CERN there are millions of collisions taking place per second. Each collision generates about 1 MB of information. The computational requirements of the four experiments that will use the LHC are enormous: each experiment will produce a few PB of data per year. For example, ATLAS and CMS foresee to produce more than 1 PB/year of raw data. ALICE foresees around 2 PB/year of raw data. LHCb will generate about 4PB/year of data.\cite{LHCDataChallenge}

All this TBs of data are generated at a single location (CERN) where the accelerator and experiments are hosted, but from that point on, innumerous activities such as digitization, reconstruction and others have to be done. The computational capacity required for those activities implies that they must be performed at geographically distributed sites. To allow that kind of execution, data and resources, must always be available to all sites in the network in a transparent and efficient way.

Besides these issues concerning data processing and resources usage, HEP impose several other requirements. One job normally involves the usage of one or more datasets. Each dataset is composed by several events and each event has its own structure. All this information must be supported by the system.\cite{Hepcal}

The solution of these issues calls for simple, efficient and reliable distributed tools.

%<<< SECTION 2 >>>
\section{BOINC and Similar Tools}

BOINC stands for Berkeley Open Infrastructure for Network Computing. It is a software platform for distributed computing developed by the same team that developed SETI@Home. It is a new framework designed to make volunteer-based distributed computing. Any computer connected to the Internet can take part in BOINC's computational efforts.

One practical example where BOINC can be used are research projects eager to use these ``almost infinite'' number of computers to increase their computing power. This is that is now called public computing. Public computing can provide more computing power than any supercomputer, cluster, or grid, and the disparity will grow over time.\cite{PublicComputing}. Current Public Computing projects can provide some indicators. For example, SETI@home run on about 1 million computers\cite{Seti}, providing a processing rate of 60 TeraFLOPs. In contrast, one large conventional supercomputer can provide about 12 TeraFLOPs. If we accept the projection that in 2015 there will be 150 million PCs connected to the Internet, then the computing power may ascend to many PetaFLOPs.\cite{PublicComputing}

\subsection{BOINC Key Concepts}

\begin{itemize}
\item \textbf{Project:} A project is a group of distributed applications, run by one organization. Projects are independent, each one has its own applications, databases and servers.
\item \textbf{Application:} This is one program dedicated to one specific computation, made up by several workunits that will produce results. It may have several versions and one application can include several files.
\item \textbf{Workunit:} One workunit describes one computation that has to be done.
\item \textbf{Result:} One result is one instance of a computation at any of its possible states.
\end{itemize}

\subsection{Features}

BOINC works in a similar way as SETI@home, its main difference being that it is able to support many other applications from within its framework. Any existing application, in common languages such as C, C++ or Fortran, can run as a BOINC application with little or no modification, only a few BOINC specific methods have to be used. Applications and associated input/output data are not physically limited since BOINC supports production/consuming of large amounts of data.

Users can run many different projects simultaneously. Currently, there are several public projects running on BOINC worldwide.

BOINC is fully manageable through its web-based system where it is possible to set up how BOINC should use the available resources. In this web-based system it is also possible to check time-varying measurements such as CPU load, network traffic and database table sizes. This simplifies the task of diagnosing current state and performance problems. 

Another feature of BOINC is fault-tolerance, since it can have separate scheduling and data servers with multiple servers of each type. Thus, if one of these servers is down another will guarantee the execution of BOINC tasks.

In terms of security BOINC is protected from several kinds of attacks. For example, to avoid the distribution of viruses it uses digital signatures based on public-key encryption. To avoid denial of services attacks, each result file has an associated maximum size. 

The implemented credit system allows to rank users and groups of users according to their computational efforts.

\subsection{Behavior}

\begin{figure}[!htb]
\centering
\includegraphics[width=65mm]{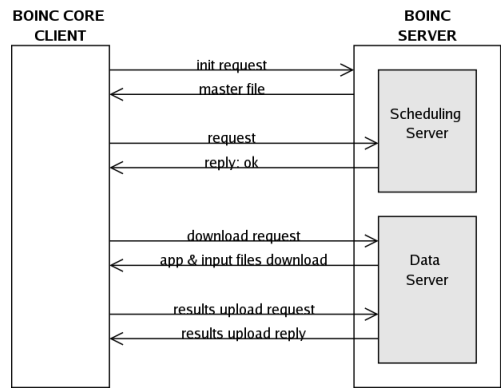}
\caption{BOINC Data Movement}
\label{boinc_data_movement}
\end{figure}

For our work it is extremely important to understand how BOINC manages data. Figure \ref{boinc_data_movement} describes this behavior. After the initial communication, the client requests work to the scheduling server. In this request the client only gives information about its hardware characteristics. According to this information, the scheduling server checks whether the client is able to run one of the available jobs. If it does, one reply is sent and the client requests to download the application and the input files. 

Then, there is a certain time limit in which the client has to compute the workunit and send the result back to the server.

\subsection{Related Work}

Nowadays, we can find an increasing number of distributed computing solutions, ranging from single volunteer-user-applications to dedicated clusters, from open source to commercial solutions, from dedicated to more generic solutions.

Since our goal is distributed computing for HEP, some projects with specific solutions using dedicated applications such as Seti@Home, Distributed.net, Folding@Home, etc, cannot be used out of the box. But the importance reached by those projects serves as a proof that their approach to computing large amounts of data is very successful.

There are some commercial applications for distributed computing such as Entropia, Data Synapse, Parabon, Avaki, and United Devices.

As a related work we can mention XtremWeb, a distributed computing tool used to generate Monte Carlo showers.\cite{XtremWeb} We can also mention JXGrid a generic distributed computing tools that can process HEP applications.\cite{JxGrid}

%<<< SECTION 3 >>>
\section{HEP@HOME}

Considering the requirements and use cases of many HEP activities and the features of BOINC, we realize that a BOINC HEP specific version could be an important and helpful tool for the physicist's daily tasks.

\subsection{Additional Features}

One of HEP@HOME main design goals is to avoid data movement. In principle, jobs run where their input data is located. This is an important issue since in HEP, input files are normally very large; thus, it avoids heavy data transfers.

In contrast to BOINC, where for a given project users only run predefined project-specific applications, in HEP@HOME users can be available for processing their own applications.

Given the fact that BOINC allows applications to have multiple files, an environment management system was defined. This allows and simplifies the usage of files associated to a certain application such as libraries, scripts, configuration files, job options files, etc. Together with the main application, these files can clearly define the conditions of a certain execution. Therefore, using these environments we have the possibility to re-execute any job. To make this mechanism even more useful, environments can be tuned by the submission of a patch. This allows users to change only the crucial aspects of one job execution. For example, the environment file of a reconstruction job contains all job options files plus several scripts. For this environment we can have two patches to make reconstruction for 10 events and 100 events.

HEP@HOME allows users to manage their own input data. When creating one workunit, besides uploading the environment/patch the user has to submit an identification of the input file he wants to work on, and a description of the result file that his work will generate. Then, his job will run in the client which has the specified file. If none of the clients has the file then the job will not run. Optionally he may submit one secondary ``get input'' application, that defines where/how the file can be found/generated. This is useful when none of the clients has the required file. In this case, the ``get input'' application will be set to run according to a predefine policy. Hence, even if he does not know whether the files he wants to work on are available in some client or not, the user has the guarantee that the computation will be done --- some client has or will have the required input file.

Normally, different HEP events are independent. Datasets are composed by events which do not have any connection among them. On the other hand, algorithms may have some sort of sequence and have to be executed according to it. HEP@HOME implements a simple mechanism to allow ordered work execution.

\subsection{Behavior}

To allow job execution according to HEP specific data movement requirements, several developments were made in BOINC components.

\begin{figure}[!htb]
\centering
\includegraphics[width=65mm]{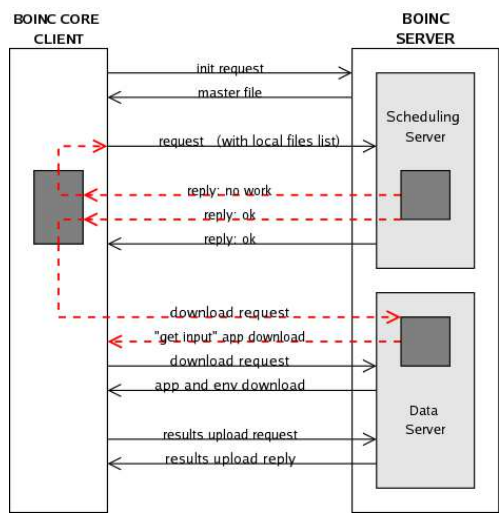}
\caption{HEP@HOME Data Movement}
\label{hephome_data_movement}
\end{figure}

In figure \ref{hephome_data_movement} we can see that after the initial communication, the client requests work to the scheduling server. In this request the client now gives information about its hardware characteristics and a list of all the available input files it has. After, the scheduling server checks if the client is able to run one of the available jobs. Two possible situations can occur for a given job:

\begin{itemize}
\item the client has the required input files. In this case an ok reply is sent, 
\item the client does not have the required files. In this case no work is sent. The server waits for a certain time according a predefined policy. This policy is based on RPC communication with the clients. At the end of this period, if none of the available clients has declared to have the necessary input files, the next client to request work can download the ``get input'' application, which will tell this client how to generate/get the input files. After this application is computed, this client will declare to have the input file it has just generate/download the next time it communicates. Server will then sent the ok reply.
\end{itemize}
The client then requests to download the application, its environment and the patch to apply. The input files are not downloaded since they are already in the client. When the computation is done the results are uploaded.

\subsection{Web interface}
BOINC's generic web interface is very complete. In order to implement the described additional features, HEP@HOME has introduced new interfaces. Although able to allow submission of several applications, only ATLAS jobs can be submitted to this web interface at this time.

%<<< SECTION 4 >>>
\section{ATLAS Use Case}

In this section we present one use case to show how can physicists use this tool to run their ATLAS jobs. This use case's actor can be the physicist doing either personal jobs submission or real production.

Let us suppose these initial facts: We have several ATLAS jobs to run, we know what each job will generate and consume and where to generate or get those files. Finally we have computers connected to the Internet ranging from simple desktop computers to cluster systems spread across the world. Any computer connected to the Internet is able to take part in this computation; the only restrictions are the job-specific requirements.

The execution process is very simple. After selecting the ATLAS application he has previously submitted, the user submits his work: the environment files (job options files, scripts, etc), a patch to apply to this environment to specify how many and what events to use, one template describing the input files and, optionally, the ``get input'' application for the input files and another template describing the result files.

As result, the user gets the aggregation of the several output files produced, in a unique output file which can be downloaded to his local computer.

%<<< SECTION 5 >>>
\section{Tests and Results}

In order to test the architecture developed and to get example results for the system behavior, several tests have been made. The defined jobs represent a complete execution of a typical sequence of physics tasks using Muon events: generation, simulation, digitization and reconstruction. All these steps were based on two main variables: $e$ - number of events and $n$ - number of CPUs running BOINC. The sequence of one execution was:

\begin{itemize}
\item 1st) Muon Generation: $e$ events (1x)
\item 2nd) Muon Simulation: $e/10$ events (10x)
\item 3rd) Muon Digitization: $e/10$ events (10x)
\item 4th) Muon Reconstruction: $e/10$ events (10x)
\end{itemize}

Two groups of tests were defined: Group A where $e=100$ and Group B where $e=1000$. For each of these groups, variable $n$ was tested with the following values: $n=1$, $n=2$, $n=8$. For each group the defined sequence was also tested directly in one computer (not using BOINC).

\begin{figure}[htb]

\centering

\includegraphics*[width=60mm]{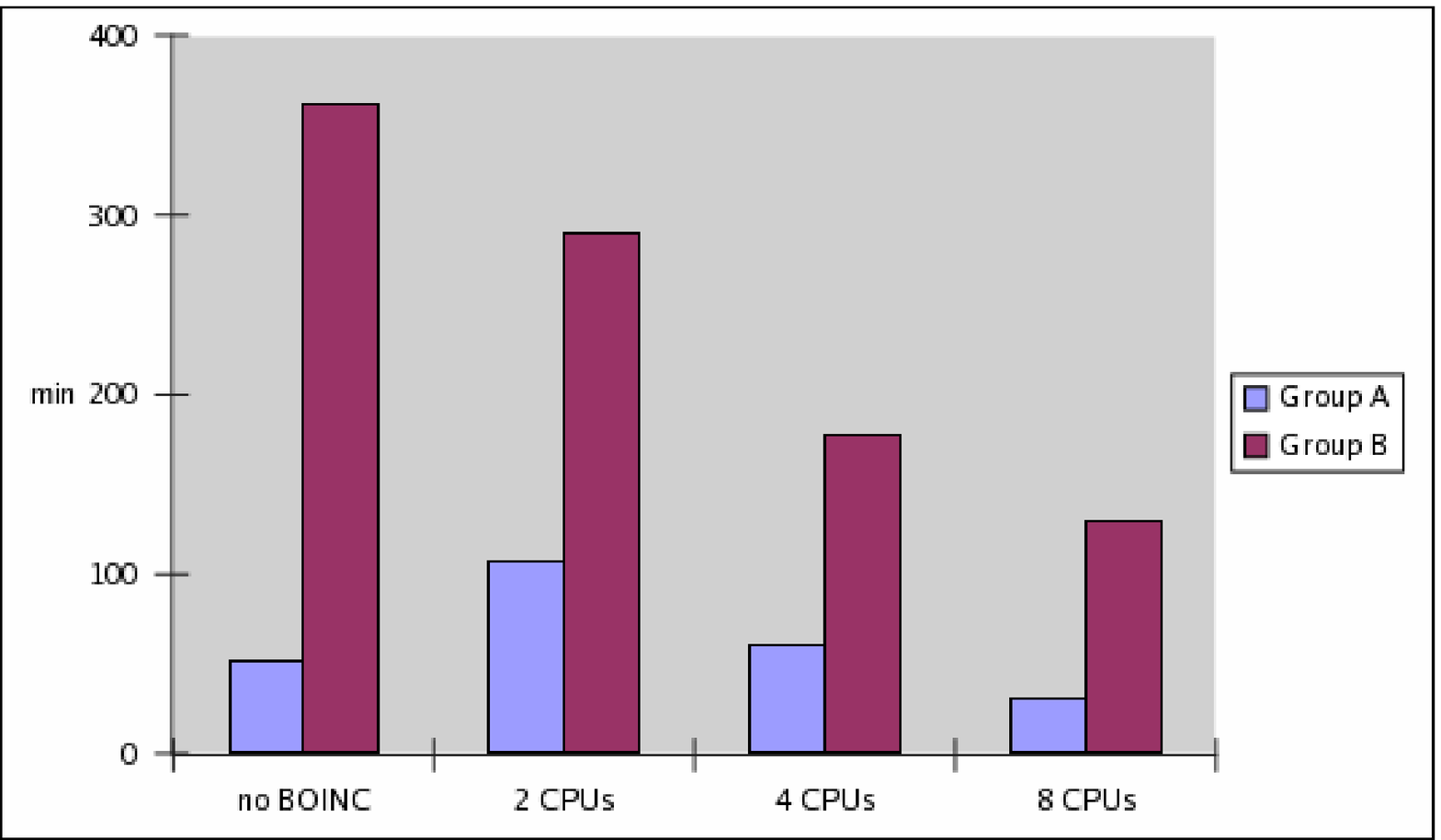}
\includegraphics*[width=60mm]{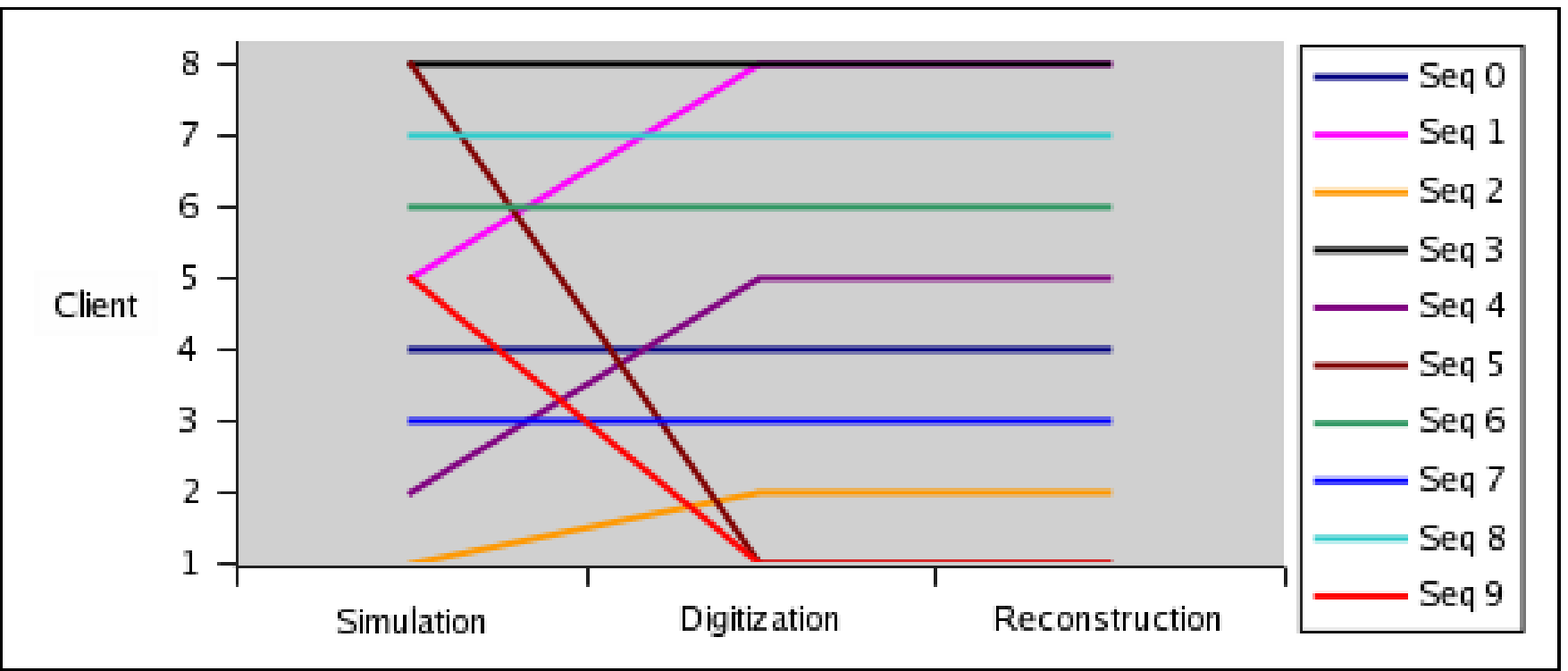}

\caption{HEP@HOME Results}
\label{graph}
\end{figure}

The columns graph in figure \ref{graph} show us the results obtained for both groups. As we can see, in group A, with 8 clients running we achieve almost half of the time of a non-BOINC execution. The execution time with two and four BOINC clients is worst than not using BOINC. These results can be explained by the overhead introduced by the communications between the BOINC server and the clients, and by the fact that in this group of tests the number of events to process is very small (only 100). 

On the other hand, in Group B (1000 events), the non BOINC execution was clearly the worst. In this case with 1000 events, the computation is heavier than with 100 events; therefore, the overhead introduced by the communications becomes less important.

In the lines graph of figure \ref{graph} we can see the information regarding data movement. In most cases execution was made where data is stored.

%<<< SECTION 6 >>>
\section{Conclusions and Future Directions}

Developing a specific tool for HEP is a complex problem since several issues related to data and resources availability have to be considered. 

Based on the success of SETI@HOME, BOINC as a generic distributed computing platform appears as a good solution to deal with that complexity. Using BOINC, our efforts were focused to HEP specific issues.

As the results show, HEP@HOME can produce faster results with no prejudice in the reliability. The tests performed have proved that the bigger the complexity of the computation (as is the case in HEP) and the bigger the number of clients, the better the improvement compared to non-distributed results. We can also conclude that we manage to avoid data movement. Finally, HEP@HOME gives physicists the possibility to submit their own jobs with the guarantee that the input data will always be available.

As future plan, one first topic to implement is to make the BOINC server decide which client to run based on its characteristics, on the presence of the input files and on the presence of the environment too. Also, our work must be focus in the optimization of several issues. The web interface can be improved allowing an easier and friendlier way to submit jobs, either ATLAS tasks or others. Special attention must also be given to the communications among server and clients avoiding inefficient communication. The optimization of clients usage avoiding idle times can be also improved.

%<<< SECTION 7 >>>
\section*{Acknowledgments}
This work was supported by the \emph{Funda\c{c}\~ao da Ci\^encia e Tecnologia} under the grant POCTI/FNU/43719/2002.


\begin{thebibliography}{9}



\bibitem{LHCDataChallenge}
Wolfgang von Rueden and Rosy Mondardini.
\newblock {The Large Hadron Collider (LHC) Data Challenge}.

\bibitem{Hepcal}
F. Carminati, P. Cerello, C. Grandi, E. Van Herwijnen, O. Smirnova, J. Templon
\newblock {Common Use Cases for a HEP Common Application Layer}.

\bibitem{PublicComputing}
David P. Anderson.
\newblock {Public Computing: Reconnecting People to Science}.
\newblock in {\em Conference on Shared Knowledge and the Web, Madrid}, 2003.

\bibitem{Seti}
David P. Anderson and Jeff Cobb and Eric Korpela and Matt Lebofsky and Dan Werthimer.
\newblock {SETI@home: an experiment in public-resource computing}.
\newblock in {\em Commun. ACM}, vol 45, number 11, pages 56--61. ACM Press, 2002.

\bibitem{XtremWeb}
Oleg Lodygensky and Alain Cordier and Gilles Fedak and Vincent Neri and Franck Cappello.
\newblock {Auger XtremWeb: Monte Carlo computation on a global computing platform}.
\newblock in {\em CHEP03, La Jolla California}, 2003.

\bibitem{JxGrid}
Daniel Templeton.
\newblock {JxGrid Application: Project JXTA in the Sun Grid Engine Context}.
\newblock in {\em SunNetwork Conference and Pavilion, San Francisco}, 2002.

\end{thebibliography}
\end{document}